\documentclass[%
 reprint,
 amsmath,amssymb,
 aps,
]{revtex4-1}

\usepackage{graphicx}
\usepackage{dcolumn}
\usepackage{bm}
\usepackage{amsthm} 

\newtheorem*{mydef}{Definition}

\newtheorem{thm}{Theorem}
\newtheorem{proposition}{Proposition}
\newtheorem{lemma}{Lemma}

\newtheorem{query}{Query}
\newtheorem{conjecture}{Conjecture}
\newtheorem{corol}{Corollary}

\newcommand{\card}[1]{\left| #1 \right|}

\DeclareMathOperator*{\argmin}{arg\,min}

\begin{document}

\preprint{APS/123-QED}

\title{A Depth-Optimal Canonical Form for Single-qubit Quantum Circuits}

\author{Alex Bocharov}
\author{Krysta M.~Svore}%
\affiliation{%
 Quantum Architectures and Computation Group\\
 Microsoft Research, Redmond, WA 98052 USA
}%




\date{\today}

\begin{abstract}
Given an arbitrary single-qubit operation,
an important task is to efficiently decompose this operation into an (exact or approximate) sequence of fault-tolerant quantum operations.
We derive a depth-optimal canonical form for single-qubit quantum circuits,
and the corresponding rules for exactly reducing an arbitrary single-qubit circuit to this canonical form.
We focus on the single-qubit universal $\{H,T\}$ basis due to its role in fault-tolerant quantum computing,
and show how our formalism might be extended to other universal bases.
We then extend our canonical representation to the family of Solovay-Kitaev decomposition algorithms,
in order to find an $\epsilon$-approximation to the single-qubit circuit in polylogarithmic time.
For a given single-qubit operation, we find significantly lower-depth $\epsilon$-approximation circuits than previous state-of-the-art implementations.
In addition, the implementation of our algorithm requires significantly fewer resources, in terms of computation memory, than previous approaches.
\end{abstract}

\maketitle


\section{Introduction}
\label{sec:intro}

Quantum algorithms assume the ability to perform any quantum operation,
however a scalable quantum computer will likely require the compilation of an arbitrary quantum operation into a discrete set of fault-tolerant operations.
Various methods of decomposing an arbitrary quantum gate into a sequence of gates drawn from a universal, discrete set are known \cite{BoykinEtAl1999},
and typically require first decomposing the operation into controlled single-qubit unitaries \cite{BarencoEtAl1995}, and then decomposing the single-qubit unitaries into a circuit of gates from a universal basis \cite{Kitaev1997,KitaevEtAl2002,Jones2012}.
%
Given that the Steane code \cite{Aliferis2006} and the surface code \cite{Fowler2009} yield high error thresholds, we choose to decompose into the basis containing the Hadamard operation ($H$) and the $\pi/8$ rotation ($T$), $\{H,T\}$, since both gates can be implemented fault-tolerantly in these codes.

Decomposing into a discrete gate set rarely results in an exactly equivalent unitary; the resulting sequence is more often an $\epsilon$-approximation to the original unitary.
In both cases, it is crucial for the quantum gate decomposition algorithm to minimize the circuit resources,
such as the circuit depth, the number of gates of a certain type, or the number of qubits.
Since the cost of implementing a non-Clifford gate fault-tolerantly is higher than in the case of a Clifford gate,
we choose to minimize the number of non-Clifford $T$ gates.
We call the corresponding cost the $T$-count of the sequence.
Our approach simultaneously minimizes circuit depth.

The Solovay-Kitaev theorem \cite{KitaevEtAl2002} states that for any $\epsilon$ and single-qubit gate $U$,
there exists a discrete approximation to $U$ with precision $\epsilon$ using $\Theta(\log^c(1/\epsilon))$ gates drawn from the universal, discrete gate set, where $c$ is a small constant.
A constructive proof of the Solovay-Kitaev theorem was shown by Dawson \emph{et al.}~\cite{DawsonNielsen2005} and gives an algorithm to find an $\epsilon$-approximation in time $O(\log^{2.71}(1/\epsilon))$.
The resulting gate sequence has depth logarithmic in precision $\epsilon$.

Optimizing a given cost, such as the $T$-count,
becomes especially important in the context of the Dawson-Nielsen algorithm \cite{DawsonNielsen2005}.
The algorithm begins with a base approximation and then proceeds recursively,
resulting in a circuit composed of $O(5^n)$ base circuits, where $n$ is the recursion depth.
The precision of the resulting circuit heavily depends on the precision of the base ``$0$-level" circuits;
if a base circuit has suboptimal cost, then this inefficiency is amplified upon composition.
In addition, the cost of a composition is often smaller than the sum of the costs of the factors (sub-additive);
a resulting circuit can often be compressed into a circuit with lower cost, even if the constituent factors are already optimal.

One technique for finding a better base circuit is given by Fowler \cite{Fowler2005}.
His algorithm uses previously computed knowledge of equivalent subcircuits to find a depth-optimal $\epsilon$-approximation to a single-qubit gate, and runs in exponential time (and much faster than brute-force search).
Our canonical form algorithm does not require costly uniqueness checks and is relatively parsimonious in the number of canonical circuits it generates.

Amy \emph{et al.}~\cite{AmyEtAl2012} describe an algorithm for decomposing an $n$-qubit unitary into an exactly equivalent
depth-optimal circuit in time $O(d\card{\mathcal{B}}^{d/2})$, where $d$ is the depth of the circuit and $\mathcal B$ is the basis.
The technique is based on a meet-in-the-middle algorithm and may be asymptotically better than Fowler's algorithm when determining exact sequences.
Their approach can also be used for multi-qubit circuit decomposition.
We note that for single-qubit circuits, our canonical form algorithm can be used to find an exact decomposition, if it exists, in improved time complexity $O(d\card{\mathcal{B}}^{d/4})$, where $\mathcal B=\{H,T\}$.

In this paper, we derive a canonical form for single-qubit unitaries.
A similar representation was given by  Matsumoto and Amano \cite{MatsumotoAmano2008}, who develop a {\it normal form\/} for $\{H,T\}$-circuits, where {\it two circuits in normal form compute the same unitary matrix if and only if the two circuits are syntactically identical} \footnote{Our work was developed independently; we were made aware of this work while writing our paper.}.
The first key difference between our canonical form and the normal form in \cite{MatsumotoAmano2008} is that their form is expressed in $SU(2)$,
which contains a non-trivial two-element center that makes the algebra sensitive to the sign of the global phase;
in contrast, our canonical representation of circuits over the $\{H,T\}$ basis is developed using group identities in the projective special unitary group $PSU(2)$.
By factoring out the global phase and working in $PSU(2)$, we are able to further compress normal circuits.
The second key difference is our concept of {\it canonical circuit\/}, which is a unique representative of a double coset of circuits with respect to the Clifford group.
It allows further compression of the depth of a circuit by writing a circuit in the canonical form $g_1.c.g_2$,
where $g_1,g_2$ are Clifford gates and $c$ is a uniquely defined canonical circuit.
Throughout, we use $.$ to represent circuit composition.

Our primary contributions are:
\begin{enumerate}
\item We present a single-qubit canonical form and corresponding rules for reducing a single-qubit circuit into the canonical form (Sec.~\ref{sec:canonical}). 
\item We develop an algorithm for finding an exact, depth-optimal decomposition of a single-qubit unitary, if it exists, else a depth-optimal $\epsilon$-approximation (Sec.~\ref{sec:decompose}). 
\item We develop an efficient storage database of canonical circuits and an efficient search procedure over the database (Sec.~\ref{sec:database}).
\item We develop an algorithm for finding an $\epsilon$-approximation to a single-qubit unitary in polylogarithmic time (Sec.~\ref{sec:SK}). 
\end{enumerate}
We begin by describing our canonical form and the corresponding reduction rules.

\section{A Canonical Form and Canonical Reduction of Circuits}
\label{sec:canonical}


We start with $PSU(2)$ representations of the Hadamard gate $H$ and the $\pi/8$-gate $T$:
\[
H=\left[
\begin{array}{cc}
i/\sqrt{2} & i/\sqrt{2} \\
i/\sqrt{2} & -i/\sqrt{2} \\
\end{array}
\right]
,
T=\left[
\begin{array}{cc}
e^{-i \pi/8} & 0 \\
0 & e^{+i \pi/8} \\
\end{array}
\right].
\]

The Phase gate $S=T^2$ and the Hadamard gate $H$ together generate a 24-element subgroup in $PSU(2)$, which is isomorphic to the to classical Coxeter group $A_3$ and isomorphic to the 4-element symmetric group $S_4$. We denote this group as $\mathcal C$.

We introduce the following two circuits, each composed of two gates, and we call these basic circuits {\it syllables}: $TH=T.H$, and $SH=S.H$.
In $PSU(2)$, syllable $TH$ is a group element of infinite order (see Sec.~4.5.3 in \cite{IkeAndMike2000}),
whereas syllable $SH$ is a group element of order 3: $SH.SH.SH = (SH)^3 =1$.

Consider the set of all circuits generated by various compositions of $TH$ and $SH$.
We note that the basis $\{TH,SH\}$ is an equivalent universal single-qubit basis to $\{H,T\}$ since the following identities hold:
\[
H=
TH(SH)^2TH;
T
= (TH)^2(SH)^2TH.
\]
Throughout, we use $\{\cdot\}$ to indicate the basis elements of a group and $\langle \cdot \rangle$ to indicate the group generated by those elements.

We further note that because $SH$ is a syllable of order 3, any circuit in $\langle TH,SH \rangle$ can be immediately reduced to one where each $SH$-dependent subsequence is either $SH$ or $(SH)^2$.
%
We also observe that any $\langle TH,SH \rangle$ circuit with $(SH)^2$ anywhere in the interior immediately collapses to an equivalent one with smaller $TH$ count.
After reducing all of the powers of $SH$ to 0, 1, or 2,
any occurrence of $(SH)^2$ in the interior of a circuit has the $TH$ syllables on both sides and thus is a part of a $TH(SH)^2TH$ pattern that collapses to $H$ upon removal of two $TH$ syllables.
Unless this residual $H$ is on the left end of the reduced circuit, it further cancels with the $H$ of the preceding $TH$ or $SH$.
Intuitively, $(SH)^2$ should not occur in a well-formed circuit.
In fact, we find that even single occurrences of $SH$ can be, in a sense, further squeezed out of the initial sequence of a circuit, leading to the notion of a {\it canonical form}.

\begin{mydef}
A non-empty circuit in $\langle TH,SH\rangle$ is said to be \emph{normalized} if it ends with $TH$ and does not explicitly contain $(SH)^2$.
A \emph{normalized} circuit is either the identity $I$ or a non-empty normalized circuit.
\end{mydef}
In other words, a normalized circuit is either the identity $I$ or follows one of the two patterns: $c.TH$ or $c.SHTH$, where $c$ is a shorter normalized circuit.

\begin{mydef}
A normalized circuit is said to be \emph{canonical} if it does not contain $SH$ earlier than the fifth syllable.
\end{mydef}
There are only six canonical circuits with fewer than six syllables: $I$, $TH$, $(TH)^2$, $(TH)^3$, $(TH)^4$, $(TH)^5$.
The shortest canonical circuit that contains the $SH$ syllable is $(TH)^4SH.TH$.

\begin{proposition}
Each $\langle H,T\rangle$ circuit $U$ can be efficiently represented as either $U=c.g$ or $U=H.c.g$, where $c$ is a normalized circuit and $g \in \mathcal{C}$.
\end{proposition}

\begin{proposition}
Each $\langle H,T\rangle$ circuit $U$ can be efficiently represented as $U=g_1.c.g_2$, where $c$ is a canonical circuit and $g_1,g_2 \in \mathcal{C}$.
\end{proposition}
Thus the right $\mathcal{C}$-coset of an arbitrary $\langle H,T \rangle$ circuit $U$ contains either $c$ or $H.c$, where $c$ is a normalized circuit that can be efficiently identified, and the double $\mathcal{C}$-coset of $U$ contains a canonical circuit that can be efficiently identified.

We now introduce the $T$-count cost and the corresponding trace level:
\begin{mydef}
The \emph{$T$-count} of a normalized circuit is the number of $TH$ syllables in that circuit.
\end{mydef}

\begin{mydef}
A \emph{trace level} $L_t$ corresponding to a value $t$, where $0\leq t \leq 2$, is the set $$L_t = \{U \in PSU(2) \Big| |tr(U)| = t\}.$$
\end{mydef}

$T$-count is an invariant of the gate represented by a canonical circuit, which follows from:
\begin{thm}\label{thm1}
If $c_1, c_2$ are $\mathcal{C}$-equivalent canonical circuits, i.e., $\exists g_1,g_2 \in \mathcal{C}$ such that  $c_2$ and $g_1.c_1.g_2$ evaluate to the same gate in $PSU(2)$,  then $c_1$ and $c_2$ are equal as $\langle TH,SH \rangle$ circuits.
\end{thm}
The proof of Theorem \ref{thm1} is given in Appendix \ref{sec:proofthm1}.

Note that in our proposed canonical form, the $T$-count and and the overall circuit depth are closely tied,
e.g., with a $\{H,T\}$ canonical form there are at least $T\mbox{-count}-1$ and at most $T\mbox{-count}+1$ Clifford gates in the representation, and all but at most two of these gates are either $H$ or $HSH$ (the number of $HSH$ sequences is guaranteed to be less than $T\mbox{-count}-3$).

\section{Depth-optimal Circuit Decomposition}
\label{sec:decompose}

A natural technique (e.g, Fowler \cite{Fowler2005}) for finding a depth-optimal $\epsilon$-approximation of $U$ is to incrementally build a database containing
unique quantum gates and their depth-optimal (shortest length) circuit representation,
and then for a given target gate $U$, perform a proximity search in the database.
Such a database of unique gates is expensive to build, store, and search.

In contrast, a database of canonical circuits can be built without recursion and requires less memory for storage,
allowing significantly longer (canonical) circuits to be maintained in practice.
The following remarkable observation leads to a more efficient algorithm (than brute-force search and \cite{Fowler2005}) for finding a depth-optimal $\epsilon$-approximation:
\begin{corol}
Given a single-qubit gate $U \in PSU(2)$,  $U$ can be $\epsilon$-approximated with an $\langle H,T \rangle$ circuit with $T\mbox{-count} < t$ if and only if one of the gates in the double coset $\mathcal{C}.U.\mathcal{C}=\{g_1.U.g_2 \Big| g_1,g_2 \in \mathcal{C}\}$ can be $\epsilon$-approximated by a canonical circuit with $T\mbox{-count} < t$.
\end{corol}
It follows that the optimal $\epsilon$-approximation of $U$ under a certain $T$-count $t$ is immediately derived from the optimal $\epsilon$-approximation of {\it some} gate $G \in \mathcal{C}.U.\mathcal{C}$
under $T$-count $t$.

The search for metric neighbors of target gate $U$, where the measure is trace distance, in a database of {\it all} unique gates is then replaced by a search for metric neighbors of all elements of the $\mathcal{C}.U.\mathcal{C}$ coset in the database of canonical circuits.
We note that there are at most $24\times 24=576$ elements in this coset and all of the searches can be done in parallel.
The design of a scalable circuit look-up solution based on the canonical representation is discussed in more detail in the next section.

Fowler has compiled the multiplication table for the group $\mathcal{C}$ generated by $H$ and $S=T^2$ (see Appendix A1 in \cite{Fowler2005});
here we use the same notation for the group elements.
The $H,S$ representations of these elements can be found in Appendix \ref{app:elts}.
Effective normalization of circuits relies on commutation relations between elements of $\mathcal{C}$ and the $T$ gate.
There are three types of relations, established by direct computation in $PSU(2)$ and catalogued in Appendix \ref{sec:relations}:
(1) $g_1.T = T,g_2$,
(2) $g_1.T=H.T.g_2$,
(3) $g_1.T=HSH.T.g_2$,
where $g_1,g_2 \in \mathcal{C}$.

In order to work constructively with normalized and canonical circuits, we prove the following propositions:
\begin{proposition}
The cost of finding a normalized representation $U=c.g$ or $U=H.c.g$ of an $\langle H,T\rangle$ circuit  $U$ is linear in the size of the circuit.
\end{proposition}
The proof of Propositions 1 and 3 is based on the actual normalization algorithm presented in Appendix \ref{sec:proof13}.

\begin{proposition}
 The cost of finding a canonical representation $g_1.c.g_2$ of an $\langle H,T\rangle$ circuit $U$ is quadratic in the $T$-count of its normalization in the worst case.
\end{proposition}
We prove Propositions 2 and 4 in Appendix \ref{sec:proof24}.

The inverse of a non-empty normalized circuit is not a normalized circuit.
However, its special form is described in the following proposition:
\begin{proposition}
Normalized representation of the inverse $c^{-1}$ of a normalized circuit $c$ is either of the from $H.c'.H$ or of the form $H.c'.H.S^3$, where $c'$ is a normalized circuit computable in time linear in the depth of $c$.
\end{proposition}
Canonical circuits are parsimonious in terms of resource requirements on a classical computer.
There are $2^{t-3}+4$ canonical circuits with $T$-count $t$ or less; for example, at $t=24$ the cardinality is $2,097,156$ and the efficient lookup tree used to experiment with circuits of this size has a memory footprint of approximately 900 MB.
A classical database of canonical circuits can be used for many practical applications, including algorithms for performing Solovay-Kitaev decomposition \cite{DawsonNielsen2005}.
We describe the classical database and how to search it efficiently in Section \ref{sec:database}.

\section{Search for Canonical Approximations}
\label{sec:database}

Let $\mathcal{B}=\{b_1, b_2, ..., b_k\} \subset PSU(2)$.
We say that $\mathcal{B}$  is a basis {\it with Clifford reduction} if there is a proper subset $\mathcal{CC}$ (to represent ``Canonical Circuits") of the subgroup $\langle \mathcal{B}\rangle$ of all of the circuits in basis $\mathcal{B}$ and a computable mapping $Cr:\langle \mathcal{B} \rangle \rightarrow \mathcal{CC}$ where $ \forall U\in \langle \mathcal{B}\rangle$, $\exists g_1,g_2 \in \mathcal{C}$ such that $U = g_1.Cr(U).g_2$.

We also assume that there is a partial function
\[cost:\langle B \rangle \rightarrow Z_{+}\]
that is
(1) well-defined on $\mathcal{CC}$;
(2) zero on $\mathcal{C}$; and
(3) subadditive w.r.t. composition, i.e., $ cost(U_1.U_2) \le cost(U_1) + cost(U_2)$ (whenever both the left-hand side and the right-hand side are well-defined).
We may additionally assume that the $cost$ function is strictly additive on $\mathcal{CC}$.

Our findings below apply to any such basis, even though the implicit focus of this section is on the $\{H,T\}$ basis with the $T$-count as the target cost function.
Consider the $\epsilon$-approximation of a target gate $U \in PSU(2)$ to precision $\epsilon > 0$.
Given a classical database of some circuits in the basis $\mathcal{B}$, the database query of primary interest is to find the minimum cost $\epsilon$-approximation of $U$:
\begin{query}
    \emph{Find} $\argmin_{v \in \langle\mathcal{B}\rangle} (cost(V) \Big| dist(V,U) < \epsilon)$.
\end{query}

Suppose now that we only have a database of some circuits in the subset $\mathcal{CC}$.
The hypothetical approximating circuit $V$ can be represented as $h_1.Cr(V).h_2, h_1,h_2 \in \mathcal{C}$.
The $cost(h_1.Cr(V).h_2) \leq cost(Cr(V))$, by the assumed properties of the $cost$ function.
We also have that $dist(h_1.Cr(V).h_2, U) = dist(CR(V) , h_1^{-1} U h_2^{-1})$.

We can now rewrite the query as
\begin{query}
    \emph{Find} $$\argmin_{g_1, g_2 \in \mathcal{C}, c \in \mathcal{CC}} (cost(c) \Big| dist(c , g_1.U.g_2) < \epsilon).$$
\end{query}

Consider the adjoint action of $\mathcal{C}$ on $PSU(2)$:
$$
    Ad_g[U] = g.U.g^{-1} , g \in \mathcal{C}, U \in PSU(2).
$$
Since $g_1.U.g_2= g_1.U.(g_2.g_1).g_1^{-1}= Ad_{g_1}[U.(g_2.g_1)]$, the query can again be rewritten as:
\begin{query}
    \emph{Find} $$\argmin_{g, h \in \mathcal{C}, c \in \mathcal{CC}} (cost(c) \Big| dist(c , Ad_g[U.h]) < \epsilon),$$
\end{query}
which is equivalent to
\begin{query}
    \emph{Find} $$\argmin_{h \in \mathcal{C}} (\min_{g \in \mathcal{C}, c \in \mathcal{CC}} (cost(c) \Big| dist(c , Ad_g[U.h]) < \epsilon).$$
\end{query}

The final query above is scalable because
the adjoint action $Ad_g$ preserves the absolute matrix trace, whereas the right action $U \rightarrow U.h$ tends to change the absolute matrix trace (for non-trivial elements of $\mathcal{C}$).
Thus the set $\{U.h \Big| h \in \mathcal{C}\}$ tends to be distributed across several (up to $|\mathcal{C}|=24$) trace levels.

We use the absolute matrix trace as the primary key in our database of $\mathcal{CC}$ circuits.
We also assume that the proximity of two circuits implies the proximity of their absolute matrix trace values.
This is obviously true when the distance measure is given by
$$
dist(U,V) = \sqrt{(2-|tr(U.V^{\dagger}|)/2},
$$
where $dist(U,V) < \epsilon$ implies that $||tr(U)| - |tr(V)||< 4 \epsilon$.
Throughout, we assume this distance measure, although other distance measures are possible.

Now consider the list of distinct absolute trace values $\{t_1, ..., t_r\} = \bigcup \{|tr|U.h| \Big| , h \in \mathcal{C}\}$ appearing in Query 4.
When $\epsilon$ is small enough, the individual approximation targets ${Ad_g[U.h], h \in \mathcal{C}}$ are distributed across non-intersecting neighborhoods
$\{U \Big| ||tr(U)| - t_i| < \delta\}$, for $i= 1, \ldots, r$ and some suitable $\delta > 0$.

Thus given that the database of the $\mathcal{CC}$ circuits is distributed across logical computational nodes indexed by the absolute trace values,
we have a good mapping of approximation target cases ${Ad_g[U.h], h \in \mathcal{C}}$ across $r$ non-intersecting logical computational node groups.

Before describing ways of further partitioning the search space, we make the following empirical observations:
\begin{enumerate}
\item Canonical circuits with $T\mbox{-count} \leq k$ have only $O(2^{k/2})$ distinct absolute trace values (empirical estimate: $ \leq 6\times2^{k/2}$ trace values).
\item Each trace level $L_t$ has either zero or at most $O(2^{k/2})$ canonical circuits with the $T$-count $k$. (Whenever Conjecture 1 of Sec.~\ref{sec:conclude} holds, the $T$-count is constant on trace level $L_t$).
\item The complexity of a search for the $\epsilon$-approximation in the database of all canonical circuits with $T$-count $\leq k$ is $O(\epsilon k 2^{k})$ when the desired approximation exists; the non-existence of the approximation is discovered in $O(k)$ steps on average and in $O(k 2^{k/2})$ steps in the worst case.
\end{enumerate}

Now we explore the geometry of an individual trace level $L_t = \{ V \Big| |tr(V)| = t\}$.
Except for the extreme values $t=0$ and $t=2$,  this trace level has the geometry of a 2-dimensional Euclidean sphere with the adjoint action of the $\mathcal{C}$ faithful and isomorphic to the action of the group of symmetries of the octahedron with vertices $(\pm 1, 0, 0), (0, \pm 1, 0),(0,0,\pm 1)$.
The trace level $L_t$, viewed as the Euclidean sphere, can be covered with 24 fundamental tiles of this action.
For instance, we can select the spherical triangle $F_0$ with vertices at $x=y=0$, $y=z=0$, and $x=y=z$, $x>0$, $z > 0$ and generate all tiles as ${Ad_g[F_0], g \in \mathcal{C}}$.
Now, consider an arbitrary fixed $h \in \mathcal{C}$ and the trace level $\{|tr(V)| = |tr(U.h)|\}$ viewed as a the tiled sphere with the $\mathcal{C}$ tiling introduced above.
For the majority of matrices $U.h$, the individual approximation targets ${Ad_g[U.h], g \in \mathcal{C}}$ are distributed across different fundamental tiles.

\begin{figure}
  \centering
  \includegraphics[width=3.25in,height=2.44in]{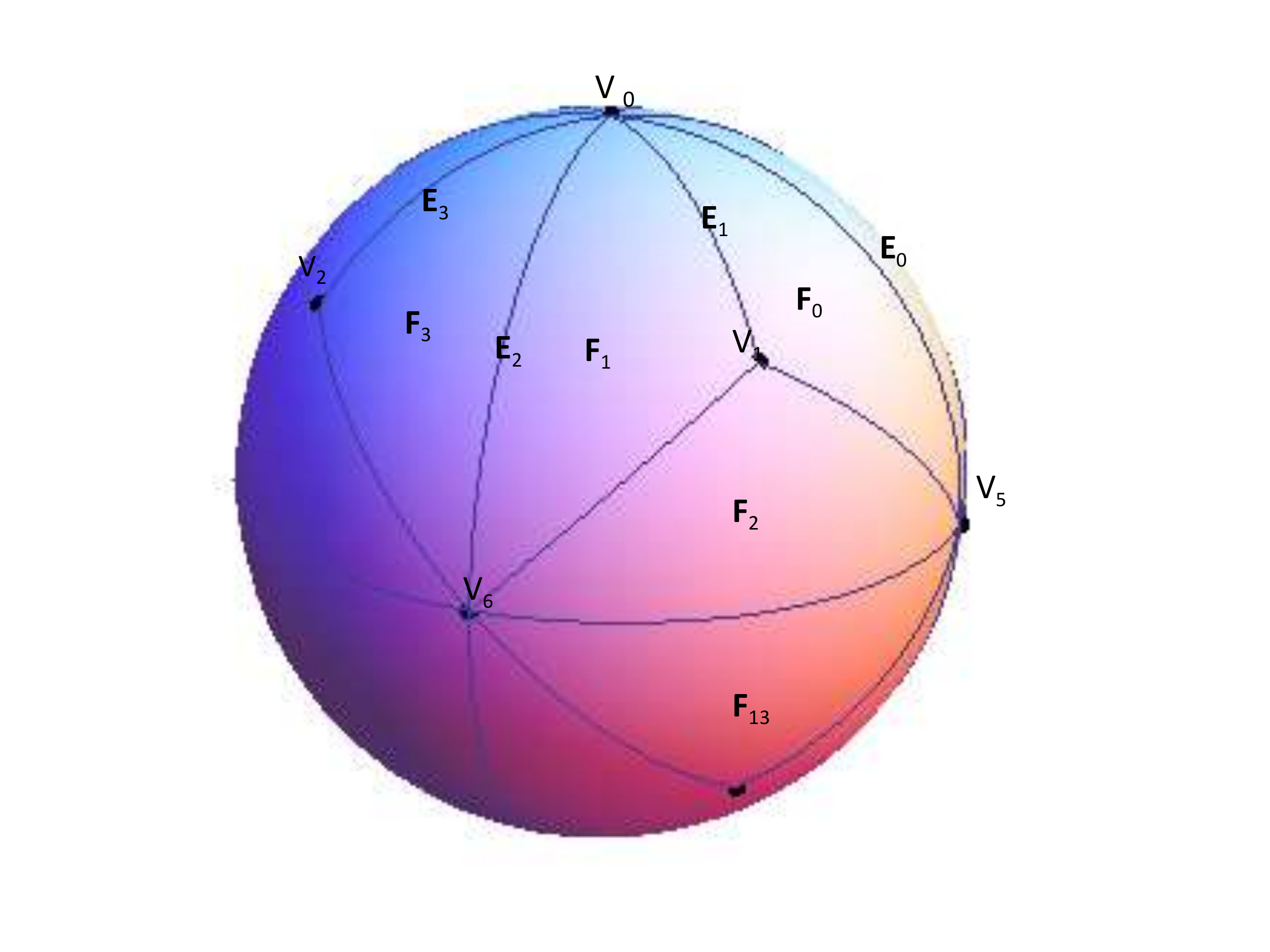}
  \caption[Trace level with 5 (out of 24) tiles and 6 (out of 14) vertices showing. $E_i, F_i, V_i$ indicate edge, face, and vertex indices, respectively.]{Trace level with 5 (out of 24) tiles and 6 (out of 14) vertices showing. $E_i, F_i, V_i$ indicate edge, face, and vertex indices, respectively.}
  \label{fig:sphere}
\end{figure}

Based on these considerations we add a collection of secondary indices to the database of the $\mathcal{CC}$ circuits where the secondary keys are provided by the geometry described above.
Given $ 0 < t < 2 $ is the value of the absolute matrix trace of certain circuits from $\mathcal{CC}$, each fundamental tile $F_i$ of the trace level $L_t$ has a {\it face index} associated with it that lists all circuits found in the interior of $F_i$.
Additionally, each pair of adjacent tiles has an {\it edge index} $E_i$ associated with it that lists all circuits for which their common boundary of is the closest such boundary.
Finally, we note 14 special points called {\it vertices} on the trace level $L_t$ that are meeting points of more than two tiles (see Figure \ref{fig:sphere}).
Each vertex $\nu$ has a {\it vertex index} $V_i$ associated with it that lists all circuits in $L_t$ for which $\nu$ is the closest vertex.

Consider the target $U.h$ of the subquery of Query 4:
$$
     \min_{g \in \mathcal{C}, c \in \mathcal{CC}} (cost(c) \Big| dist(c , Ad_g[U.h])) < \epsilon.
$$
Let $0 < t < 2$, such that $||tr(U.h)| - t|< 4 \epsilon$ and trace level $L_t$ contains some circuits from $\mathcal{CC}$.
For the majority of matrices $U$, the projection of $U.h$ on the trace level $L_t$, with high probability, is far enough from boundaries of the fundamental tile $F$ to the interior of which that projection belongs.
Therefore in order to find the  $\min_{c \in \mathcal{CC}} (cost(c) | dist(c, U.h)) < \epsilon)$ in this case it suffices to inspect the face index of that tile. For a non-trivial $g \in \mathcal{C}$ the situation is isometric, so the search for $\min_{c \in \mathcal{CC}} (cost(c) | dist(c, Ad_g[U.h])) < \epsilon)$ can be limited to the interior of the $Ad_g[F]$ tile.

Of course with lower probability, $U.h$ will fall within $\epsilon$ of some edge or vertex of the trace level $L_t$,
which requires the use of multiple tile, edge or vertex indices.
In practice the above subquery should be distributed over all relevant secondary indices.
With high probability, most of the indices will be immediately eliminated based on the trace-level geometry.

\section{Application to Solovay-Kitaev Decomposition}
\label{sec:SK}

In this section, we use our canonical representations for Solovay-Kitaev decomposition.
Recall that the Dawson-Nielsen (D-N) algorithm for the Solovaty-Kitaev theorem \cite{DawsonNielsen2005} is recursive, and finer approximations require greater recursion depth.
At depth level 0, D-N returns an extrinsic ``basic" approximation of a requested single-qubit gate $U$.
At depth $n$, it composes an approximation from the depth $n-1$ approximation $U_{n-1}$ and the depth $n-1$ approximations of two auxiliary matrices $V_{n-1}$ and $W_{n-1}$, such that
the resulting approximation is given by
\begin{equation}
U_n=V_{n-1}.W_{n-1}.V_{n-1}^{\dagger} .W_{n-1}^{\dagger}.U_{n-1}.
\label{eq:DN}
\end{equation}

We want to maintain the canonical form for each of the approximating circuits at each depth level, starting with base level $n=0$.
We can efficiently lookup the $0$-level approximations by using our design for efficient parallel lookup over a large database of canonical circuits (see Section \ref{sec:database}).
This results in an interesting tradeoff.
When all 0-level approximations are sought in a database of canonical circuits with $T$-count $\leq t$, where $t$ is relatively large,
in the worst case the D-N $n$-level recursion may result in a circuit with $T$-count cost $O(t5^n)$,
seemingly worsening the $T$-count vs.~precision performance curve for the algorithm.

On the other hand, improving the quality of the $0$-level approximation may in fact decrease the required recursion depth and exponentially decrease the circuit's $T$-count.
For example, increasing the $0$-level database scope from $T$-count $\leq 12$ to $T$-count $\leq 28$ improves the precision of the $0$-level approximation by a factor of $9.8$ on average.
According to the D-N estimate (Sec.~3, Eq 1 in \cite{DawsonNielsen2005}),
this results in an improvement in precision by a coefficient around $10^{-6}$ at depth 4 and around $10^{-9}$ at depth 6.
Thus if we have an $\epsilon$-approximation using a database containing circuits with $T$-count $\leq 12$,
then we can expect to have a significantly more precise $\epsilon$-approximation by expanding the database to include circuits with $T$-counts in the high 20's.

In practice, we find that our technique scales even better than the D-N estimate suggests.
With a database of $0$-level approximations up to $T\mbox{-count}=25$ or $26$,
we are limited as early as recursion depth 4 only by the accuracy of the machine-defined {\tt double} type.
Therefore, our experimental results only cover recursion depths $\leq 3$ \footnote{Finer analysis would seem to require extended floating point precision.}.
In terms of circuit cost, we barely exceed a $T$-count of 3000 for the longest of our circuit approximations, whereas previous approaches cite $T$-counts of $10^5$ or more.

The impact of the canonical reduction on the quality of the D-N commutant formula (Eq \ref{eq:DN}) is profound.
Consider first the composition of a canonical presentation with a normalized presentation (in this order).
Without loss of generality, we can consider composition in the form $U=(g_1.V.TH.g_2).[H.].W.g_3)$, where $g_1, g_2, g_3 \in \mathcal{C}$, $W$ is normalized, and $V.TH$ is canonical.
The $[\cdot]$ indicates that the sequence is present in one case and absent in the other.
We are especially interested in cases where cancelation occurs, namely the resulting composition has $T$-count smaller than the sum of the $T$-counts of $V.TH$ and $W$.
Cancelation is triggered by a certain structure of the normalization of the $(H.g_2.[H.].W)$ circuit that is of the form $W'=[H.][SH.]W_1.g_4$, where $g_4 \in \mathcal{C}$
and the normalized circuit $W_1$ is either empty or starts and ends with $TH$.
By Lemma 1, the trailing $T$ in $g_1.V.T$ will not cancel when $W'$ starts with $H$ or $SH$, or when $W_1$ is empty.
Consider the remaining case: $W'=TH.W_2.g_4$.
Here, $U=g_1.V.SH.W_2.g_4$, implying that $T\mbox{-count}(U) < T\mbox{-count}(V)+T\mbox{-count}(W)$.

Further transformations are necessary when $V= V_2.SH$.
If $W_2$ starts with $SH$, i.e., $W_2 = SH.W_3$, then $U=g_1.V_2.W_3.g_4$ is a normalized form and no further reduction in $T$-count is possible.
However, if $W_2$ starts with $TH$ we get the infamous $TH.(SH)^2.TH$ pattern, which reduces to $H$, which is likely to cascade into further cancelations.

To summarize, normalized composition of circuits reduces the $T$-count of the resulting circuit in many cases.
An additional benefit is that by using canonical reduction, we can restrict the number of Clifford gates as well.
Each interior gate in a normalized circuit is either $H$ or $HSH$ (and if the circuit is canonical then the number of $HSH$ gates cannot be greater than $T\mbox{-count} - 5$).

Given an $\epsilon$-approximation circuit $c$ of a target gate $U$, for example by using D-N, the normalized form of circuit $c$, denoted by $n(c)$, is a minimal cost circuit that is {\it exactly} equivalent to $c$; however, normalization does not guarantee that the result is a lowest cost $\epsilon$-approximation of $U$.
Indeed, there are potentially many normalized circuits in the $\epsilon$-neighborhood of $U$,
including some with $T$-counts lower than the $T$-count of $n(c)$, that are simply not obtainable by a specific method (e.g., the D-N algorithm for Solovay-Kitaev).

\section{Experimental Results}
\label{sec:results}

We evaluate the performance of our canonical form and reduction techniques in two experimental scenarios.
In each case, we evaluate the performance of decomposing $10,000$ randomly generated, single-qubit unitaries into their $\epsilon$-approximations.
First, we study the tradeoffs between $T$-count cost and precision $\epsilon$ for the $0$-level $\epsilon$-approximation, employing our canonical circuit database.
Second, we study the same tradeoffs for the $n$-level $\epsilon$-approximation, where $n\leq3$, using our database, canonical reduction, and the recursive Solovay-Kitaev algorithm \cite{DawsonNielsen2005}.

To evaluate our findings, we generated and catalogued each of the $268,435,460$ canonical circuits with $T\mbox{-count}\leq 31$.
Our database of canonical circuits has the absolute matrix trace as its primary index,
and has secondary indices based on the fundamental tiles of the adjoint representation of the $\mathcal{C}$ group (see Sec.~\ref{sec:database}).

Our experiments and database required a memory footprint of $120$GB and the use of a high-performance multi-core workstation.
We discovered, however, that canonical circuits with $T\mbox{-count} > 25$ did not offer significant improvements in $T$-count/precision $\epsilon$ tradeoffs in the second experimental scenario using machine {\tt double} accuracy.
In practice, a database of canonical circuits of $T\mbox{-count}\leq 25$, which has cardinality $4,194,308$ and RAM footprint $\sim$2GB, is sufficient.
In all cases, extensive multithreading is required when high query throughput is sought.

\begin{figure}
  \centering
  \includegraphics[width=3.25in,height=2.44in]{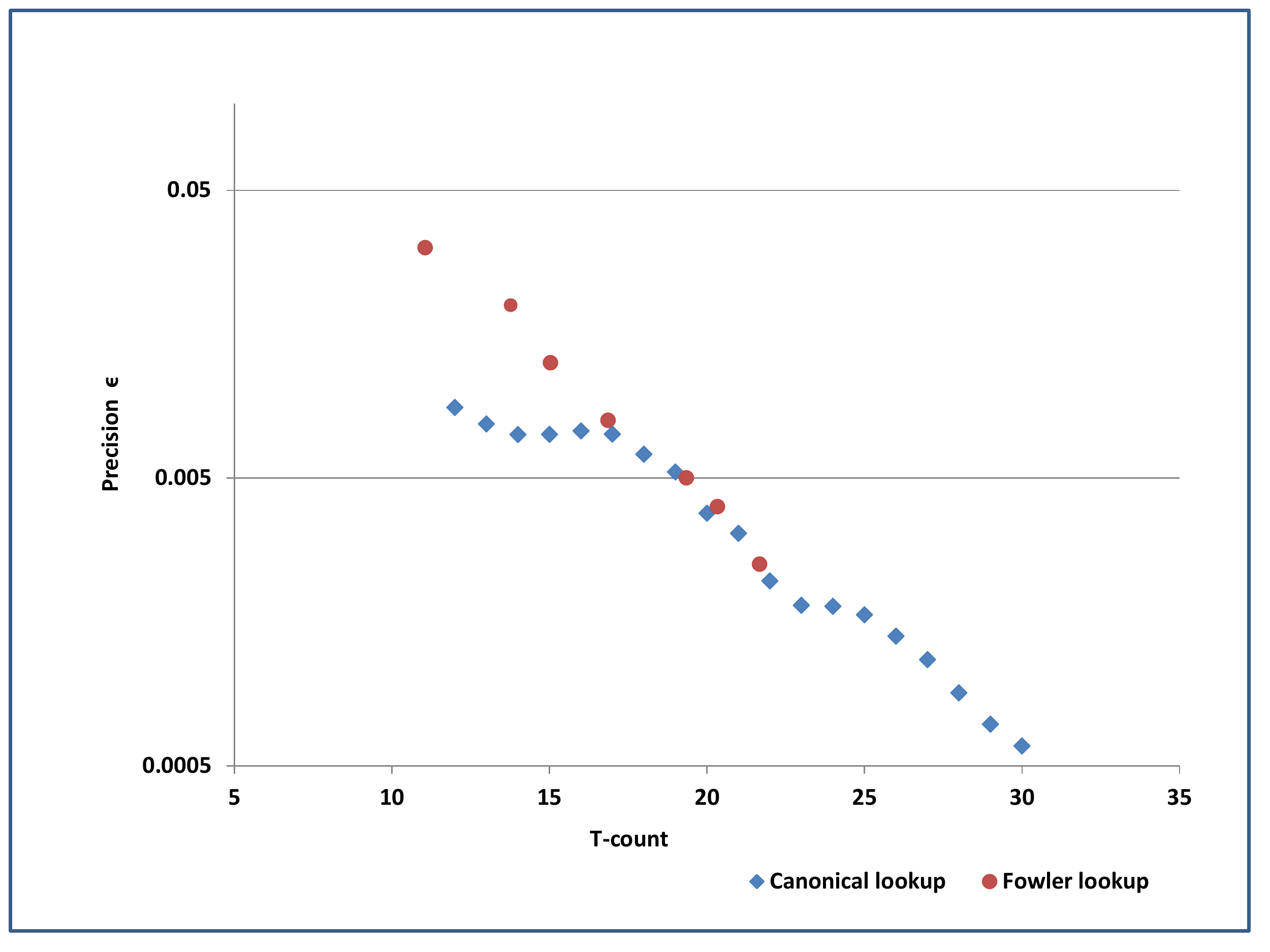}
  \caption[Canonical lookup vs. Fowler lookup]{$T$-count versus mean precision $\epsilon$ (trace distance) over the $\epsilon$-approximations at $0$-level for $10,000$ random unitaries.}
\label{fig:0LevelResults}
\end{figure}

We compare the performance of our depth-optimal $0$-level $\epsilon$-approximation invoking our canonical circuit database with the state-of-the-art, depth-optimal baseline technique of Fowler \cite{Fowler2005}.
Figure \ref{fig:0LevelResults} shows the $T$-count versus the precision $\epsilon$ for our canonical form technique (search in our database) and for Fowler's technique,
where Fowler uses a database of unique $\langle H,T\rangle$ gates.
Both curves are obtained by calculating the mean precision $\epsilon$ for a given $T$-count for the $\epsilon$-approximations of $10,000$ random unitary gates.

Since both techniques are depth-optimal, we expect the curves to align, and hope to find that our database can store much longer sequences than previous techniques.
The curves are sufficiently identical for $T$-counts between 15 and 22.
The slight divergence below $T$-count 15 is likely due to the fact that Fowler's technique optimizes for overall gate count (circuit length), whereas we optimize for $T$-count.
Fowler's method could however be adapted to minimize $T$-count, in which case the curves would be identical up to $T$-count 22.
The key observation is that reduction to canonical circuits enables a much larger database to beyond a $T$-count of $30$ (without the use of overly extravagant hardware),
where as previous state-of-the-art techniques obtain less compression, and in turn require more memory, limiting the database to $T$-count 22 \footnote{Note that each increase by 1 in $T$-count requires roughly twice the amount of memory.}.

\begin{figure}
  \centering
  \includegraphics[width=3.25in,height=2.44in]{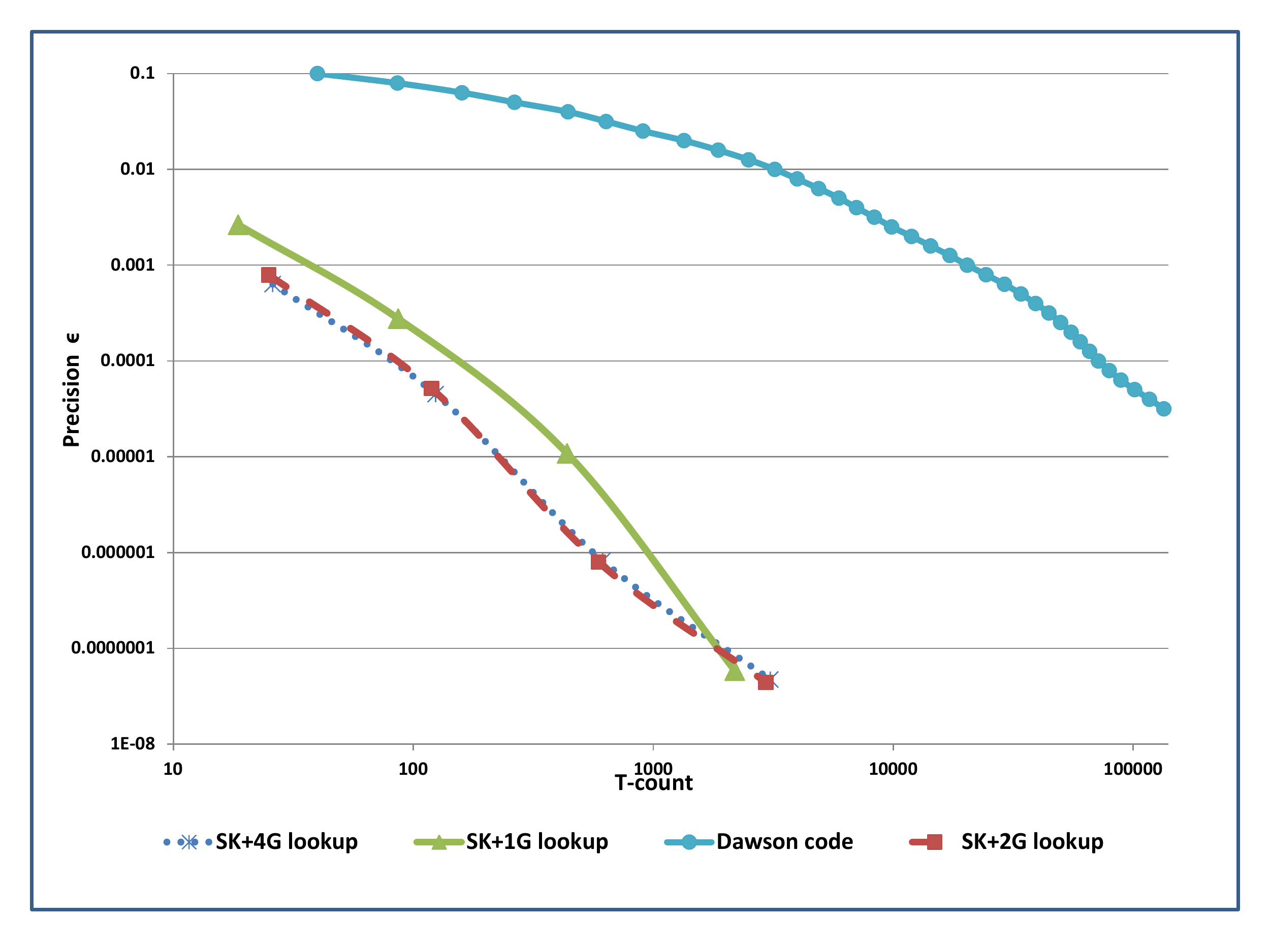}
  \caption[Solovay-Kitaev Precision Comparison]{$T$-count versus mean precision $\epsilon$ (trace distance) over the $\epsilon$-approximations at $n$-level recursion for $10,000$ random unitaries and $n=0,1,2,3$, where the markers indicate the recursion level $n$.}
\label{fig:SKResults}
\end{figure}

We next study canonical forms within Solovay-Kitaev decomposition.
We compare the use of our canonical reduction within Dawson-Nielsen's algorithm to the original Dawson-Nielsen algorithm \cite{DawsonNielsen2005}.
Figure \ref{fig:SKResults} compares three implementations of our canonical technique to D-N.
The canonical implementations use canonical reduction, as well as three different canonical circuit database sizes, 1GB, 2GB, and 4GB, each enabling storage of circuits with up to $T$-count 24, 25, and 26, respectively.
Each curve represents the mean precision $\epsilon$ for a given $T$-count for the $\epsilon$-approximations of $10,000$ random unitary gates for recursion levels $n=0,1,2,3$.
Both axes in the graph are plotted on the logarithmic scale.

First, we note that there is no visible difference between the 2GB canonical implementation and the 4GB canonical implementation.
Second, we observe that our technique, for all three implementations, is able to find, for a given $\epsilon$, approximations with significantly smaller $T$-count.
In particular, at $T$-counts below 500, our methods achieve $\epsilon = 5\times10^{-8}$, offering a factor of $10^{-6}$ improvement over D-N.
To improve the precision of our technique even further, it would require computation of the matrix trace using precision beyond the limit of machine {\tt double} precision.
At the best D-N precision of $\epsilon=5\times10^{-5}$, D-N requires roughly $100,000$ $T$ gates on average,
while our 2GB implementation (SK+2G) requires only 120 $T$ gates on average (a factor of $846$ improvement).

\section{Conclusions and Future Work}
\label{sec:conclude}

We have defined a depth-optimal canonical form and corresponding reduction rules for single-qubit quantum circuits.
Our techniques result in significant improvements in terms of database size and achieved precision in the case of the depth-optimal $0$-level $\epsilon$-approximation,
and significant improvements in the $T$-count/precision $\epsilon$ curve when applied to Solovay-Kitaev decomposition for $n$-levels of recursion.
A natural future direction is to generalize the definition of a canonical form to multi-qubit gates as well as to other universal bases.

Another direction is to perform ``lossy compression", where the task is to find an approximately equivalent circuit (within distance $\epsilon$ of the target gate) that requires less cost,
in terms of a given cost function such as $T$-count or number of gates.
We believe such a solution it will require the following conjecture:
\begin{conjecture}
If $c_1, c_2$ are canonical circuits and $T\mbox{-count}(c_1) \neq T\mbox{-count}(c_2)$ then $|tr(c_1)| \neq |tr(c_2)|$.
\end{conjecture}
This conjecture implies that if a trace level $L_t = \{U \in PSU(2) \Big| |tr(U)| = t\}$  contains multiple canonical circuits, all of these circuits have the same $T$-count.
We currently have only empirical brute-force evidence of Conjecture 1 for $T\mbox{-count}\leq 31$.

\begin{acknowledgments}
We thank Dave Wecker, Burton Smith, Michael Freedman, Zhenghan Wang and John Platt for useful discussions. We also wish to thank Rodney Van Meter and Nathan Cody Jones for sharing the benchmark D-N algorithm data with us.
\end{acknowledgments}


\bibliography{BocharovSvore_PRL}
\bibliographystyle{apsrev}

\appendix
\section{Elements of the $\mathcal{C}$ group}
\label{app:elts}

The following definitions are equivalent to the ones given in (Appendix A1 in \cite{Fowler2005}).
\[
\begin{matrix}
G_0=Id; G_1=H; G_2=HSSH; G_3=SS; G_4=S; \\
G_5=SSS; G_6=HSS; G_7=SSH; G_8=SH; \\
G_9=SSSH;G_{10}=SSHSSH; G_{11}=SHSSH; \\
G_{12}=SSSHSSH; G_{13}=HS; G_{14}=HSSS; \\
G_{15}=SSHSS; G_{16}=SHSS; G_{17}=SSSHSS; \\
G_{18}=HSH;G_{19}=HSSSH; G_{20}=HSHSSH; \\
G_{21}=HSSSHSSH; G_{22}=SSSHS; G_{23}=SHSSS
\end{matrix}
\]

\section{$\mathcal{C}/T$ commutation relations}
\label{sec:relations}

\[
\begin{matrix}
G_1.T=H.T; G_2.T=T.G_{12}; G_3.T=T.G_3; \\
G_4.T=T.G_4; G_5.T=T.G_5; G_6.T=H.T.G_3; \\
G_7.T=H.T.G_{12}; G_8.T=H.SH.T.G_2; \\
G_9.T=H.SH.T.G_4; G_{10}.T=T.G_{11}; \\
G_{11}.T=T.G_2; G_{12}.T=T.G_{10}; \\
G_{13}.T=H.T.G_4;G_{14}.T=H.T.G_5; \\
G_{15}.T=H.T.G_{11}; G_{16}.T=H.SH.T.G_{10}; \\
G_{17}.T=H.SH.T.G_{5}; G_{18}.T=H.SH.T; \\
G_{19}.T=H.SH.T.G_{12};G_{20}.T=H.T.G_{2}; \\
G_{21}.T=H.T.G_{10}; G_{22}.T=H.SH.T.G_{3}; \\
G_{23}.T=H.SH.T.G_{11}
\end{matrix}
\]

\section{Proof of Propositions 1 and 3}
\label{sec:proof13}

Since $T^2=S \in \mathcal{C}$, any  $\langle H,T\rangle$ circuit has the form
$U = (\prod_{i=1}^k g_i.T).g, k \ge 0$, where $g,g_i \in \mathcal{C}$, $g_i \neq Id$ when $i>1$.

Collect all factors in this product (in the order they appear) into a $gateList$.
The following algorithm is tail-recursive, and group $\mathcal C$ is denoted by {\tt C}:

Algorithm
\begin{verbatim}
CircuitNormalize(input: gateList):gateList =
  if gateList is empty then
    return  empty list
  let left <- {head(input)}
  let right <-  tail(input)
  while (left is not empty) &&
        (right is not empty) do
    if head(right) = T then
      if head(left) = T then
        left <- tail(left)
        right <- {G4} + tail(right)
        // G4=S=T.T
      else // head(left) in C
        if head(left) = H[SH] then
          left <- {T} + left
          right <- tail(right)
        else
          let cmt <- //see Appendix 2
            apply C/T commutation
            table to head(left) and T
          left <- tail(left)
            if (cmt = H[SH].T.g , g in C)
            then
              left <- { g, T, H[SH]} + left
              right <- tail(right)
            else if (cmt = T.g , g in C)
            then
              right <- { T, g} + right
    else
      if head(left) = T then
        left <-{head(right)} + left
      else // head(left), head(right) in C
        let g <- C product of
                 head(left) and head(right)
        left <- tail(left)
        if g <> Id then
          left = {g} + left
      right <- tail(right)
    if left is empty then
        return CircuitNormalize(right)
    else
        return reverse(left) + right
\end{verbatim}
The intent of this algorithm is to eliminate all of the Clifford gates that are different from either $H$ or $HSH$ from the interior of the ``gateList". The cost of each such elimination is bound by a constant. Thus the cost of the algorithm is linear in terms of the number of such Clifford gates and hence linear in terms of the length of the input circuit.

\section{Proof of Propositions 2 and 4}
\label{sec:proof24}

\begin{lemma}
 A normalized circuit of the form  $U=SHTH.c$ (where $c$ is a normalized subcircuit) can be effectively rewritten as a normalized representation $H.SHTH.c_1.g, g \in \mathcal{C}$ with the number of rewrites linear in the $T$-count of $c$. The resulting circuit $c_1$ has the same $T$-count as $c$.
\end{lemma}
\begin{proof}
By brute force, we establish that $SHTH=HSHT.HSS$ and ``upset" the normalization to start with $HSHT.HSS.c$. The rest of the proof is similar to the proof of Propositions 1 and 3, i.e., we establish by linear induction that $HSS.c$ reduces to $H.c_1.g, g \in \mathcal{C}$, where $c_1$ is a normalized circuit.
\end{proof}
Informally, if a normalized circuit starts with $SH$ then we can force it into a normalized presentation that starts with $H$.

We are now ready to prove Propositions 2 and 4.

\begin{proof}
Let $U=[H.]c.g$ be a normalized representation of a given $U \in PSU(2)$.
Note that $c$ may start with the $SH$ syllable, in which case, we split it off.
Now consider $U=[H.][SH.]c.g$, where $c$ is a normalized circuits starting with the $TH$ syllable.
Further proof is based on the following identities that can be established by brute-force calculation in $PSU(2)$:
\[
\begin{matrix}
THSHT=G_2.THT.G_4; THTHSHT=G_3.THTHT.G_2; \\
THSHTHT=G_{10}.THTHT.G_{11}; \\
THSHTHSHT=G_2.THTHT.G_5; \\
THTHTHSHT=G_{11}.THTHTHT.G_4; \\
THTHSHTHT=G_5.THTHTHT.G_{11}; \\
THSHTHTHT=G_4.THTHTHT.G_{12}; \\
THTHSHTHSHT=G_3.THTHTHT.G_5; \\
THSHTHSHTHT=G_5.THTHTHT.G_3; \\
THSHTHTHSHT=G_{10}.THTHTHT.G_{10}; \\
THSHTHSHTHSHT=G_2.THTHTHT.G_2;
\end{matrix}
\]

Informally, these are used to ``squeeze" $SH$ syllables out of the first four syllables of $c$ into surrounding $\mathcal{C}$ factors.
If $c$ has fewer than five $TH$ syllables, we immediately obtain $U=g_1.c'.g_2, g_1,g_2 \in \mathcal{C}$,
where $c'$ is a canonical circuit.
We now assume that $c$ has $T$-count $t > 4$ and that the propositions have been proven for all $T$-counts smaller than $t$.
Consider the shortest prefix of the circuit $c$ spanned by its leftmost four $TH$ syllables and apply one of the above transformation rules to that prefix, thus obtaining reduction of the form $U=g_1.THTHTHT.g'.c'.g, g_1, g', g \in \mathcal{C}$, where $c'$ is a normalized circuit.
Apply Proposition 1 to subcircuit $g'.c'.g$ to obtain a normalized presentation $V=[H.][SH.]c''.g'', g'' \in \mathcal{C}$, where $c''$ is a normalized circuit that is either empty or starts with $TH$. In the empty case $c''$, we trivially get the canonical presentation $U=g_1.THTHTHTH.(H.[H.][SH.]g'')$.
Otherwise, we need to consider the following three cases:
\begin{enumerate}
\item $V$ starts with $H$. This yields canonical presentation $U=g_1.THTHTHTH.[SH.]c''.g''$;
\item $V$ starts with $SH$, as per Lemma 1 we can force it to start with $H$ and reduce to the first case.
\item $V$ starts with $TH$, i.e., $V=TH.c'''.g''$, hence $U=g_1.THTHTHT.TH.c'''.g''=g_1.THTHT.HSH.c'''.g''$, where $THTHTHSH.c'''$ is normalized with $T$-count smaller than $t$. The latter is not canonical, since there is the $SH$ occurring earlier than the fifth syllable, however the circuit is normalized with $T$-count smaller than $t$ and can be recursively brought to canonical form as per the induction hypothesis.
\end{enumerate}

Note that the last case is the only one responsible for the potentially quadratic cost of the canonical reduction. Normalization of subcircuits of the above $g'.c'.g$ form has linear cost. For the overall cost to become quadratic, the circuit shape as in clause 3 must occur $O(t)$ times in the at most $t/2$ recurring rewrites, which is fairly unlikely.
In fact, in practice we have never seen clause 3 invoked in our experiments.
\end{proof}

\section{Proof of Theorem 1}
\label{sec:proofthm1}

We outline a proof by induction of Theorem 1.
It is reminiscent of Sec.~4.2 in \cite{MatsumotoAmano2008}, albeit dramatically simpler and shorter.

\begin{proof}
The simple initial step is to note that if there exist such $g_1,g_2, c_1, c_2$ that $c_2=g_1.c_1.g_2$ as matrices and $c_2 \neq c_1$ as circuits then there exists a normalized circuit $n$, with $T\mbox{-count}(n) > 0$, that evaluates to a matrix in $\mathcal{C}$.
Since $SH \in \mathcal{C}$ and $T\mbox{-count}(SH)=0$, $n$, without loss of generality, starts with $TH$.

Now consider the adjoint action of $PSU(2)$ on its Lie algebra $L=su(2)$, $ad_u[m]=u.m.u^{\dagger}$, $u \in PSU(2)$, $m \in L$.
It is a well known fact that $L$ consists of zero-trace Hermitian matrices and is spanned over $R$ by the Pauli matrices $X,Y,Z$.

The adjoint action of the $\mathcal{C}$ subgroup on $L$ is the symmetry group of the octahedron with vertices at $\pm X, \pm Y, \pm Z$. In particular, for each  $g \in \mathcal{C}$, $ad_g[Z]$ must be one of these vertices.
To obtain a contradiction it suffices to show that for a normalized circuit $n$, $ad_n(Z)$ cannot be in $\{\pm X, \pm Y, \pm Z\}$.

Let $A \in L$ be a matrix over $Q(\sqrt{2})$ represented as:
$$(\sqrt{2})^l A = (x_0+x_1 \sqrt{2}) X + (y_0+y_1 \sqrt{2}) Y + (z_0+z_1 \sqrt{2}) Z,$$
where $x_0,x_1,y_0,y_1,z_0,z_1$ are integers.

We show that if $A=ad_n(Z)$ then (1) $x_0$ is odd and (2) $y_0,z_0$ have the opposite parity.
The (1) implies that the coefficient at $X$ is non-zero and the (2) implies that at least one other coefficient (at $Y$ or at $Z$) is non-zero; together they imply that $ad_n(Z)$ cannot be proportional to any one Pauli matrix.

We prove the desired properties (1) and (2) by induction on the $T$-count of $n$.
By direct computation:
\begin{eqnarray*}
ad_{TH}(X) &=& Z, \\ad_{TH}(Y) &=& (X-Y)/\sqrt{2}, \\ad_{TH}(Z) &=& (X+Y)/\sqrt{2},\\
ad_{SHTH}(X) &=& Y, \\ad_{SHTH}(Y) &=& (-X+Z)/\sqrt{2}, \\ ad_{SHTH}(Z) &=& (X+Z)/\sqrt{2},
\end{eqnarray*}
and, in particular, properties (1) and (2) hold for $ad_{TH}(Z) = (X+Y)/\sqrt{2}$  ( $x_0=1, y_0=1, z_0=0$ ).

Given matrix $A \in L$ presented as shown above, we have:
\begin{eqnarray*}
(\sqrt{2})^{l+1} ad_{TH}(A) = ((y_0+z_0)+(y1+z_1) \sqrt{2}) X + \\ ((z_0-y_0)+(z_1-y_1) \sqrt{2}) Y + (2 x_1 + x_0 \sqrt{2})Z,\\
(\sqrt{2})^{l+1} ad_{SHTH}(A) = ((z_0-y_0)+(z_1-y_1) \sqrt{2}) X  + \\ (2 x_1 + x_0 \sqrt{2}) Y + ((y_0+z_0)+(y_1+z_1) \sqrt{2}) Z.
\end{eqnarray*}

By induction hypothesis, $y_0, z_0$ have opposite parity, therefore the new $x_0$ that is equal to either $y_0+z_0$ or $z_0-y_0$ is odd in both cases.
In the expression for  $ad_{TH}(A)$, the new $y'_0 = z_0-y_0$ is odd but the new $z'_0= 2 x_1$ is even.
In the expression for $ad_{SHTH}(A)$, the new $y'_0 = 2 x_1$ is even but the new $z'_0= y_0+z_0$ is odd.

Since each non-trivial normalized circuit is either $n_1.TH$ or $n_1.SHTH$, where $n_1$ is a shorter normalized circuit, this concludes the inductive proof.
\end{proof}

\end{document}